\documentclass[aps,pra,twocolumn,tightenlines,floatfix,showpacs,superscriptaddress,10pt, longbibliography]{revtex4-1}

\usepackage{amsmath}
\usepackage{amsfonts}
\usepackage{amssymb}
\usepackage{graphicx}

\begin{document}

\title{Instantons in self-organizing logic gates}

\author{Sean R. B. Bearden}
\affiliation{Department of Physics, University of California, San Diego, La Jolla, CA 92093}

\author{Haik Manukian}
\affiliation{Department of Physics, University of California, San Diego, La Jolla, CA 92093}

\author{Fabio L. Traversa}
\affiliation{MemComputing, Inc., San Diego, CA, 92130 CA}

\author{Massimiliano Di Ventra}
\email{email: diventra@physics.ucsd.edu}
\affiliation{Department of Physics, University of California, San Diego, La Jolla, CA 92093}




\date{\today}

\begin{abstract}
Self-organizing logic is a recently-suggested framework that allows the solution of Boolean truth tables ``in reverse,'' i.e., it is able to satisfy the  logical proposition of gates {\it regardless} to which terminal(s) the truth value is assigned (``terminal-agnostic logic''). It can be realized if time non-locality (memory) is present. A practical realization of self-organizing logic gates (SOLGs) can be done by combining circuit elements with and without memory. By employing one such 
realization, we show, numerically, that SOLGs exploit elementary instantons to reach equilibrium points. Instantons are classical trajectories of the non-linear equations of motion describing SOLGs, and connect topologically distinct critical points in the phase space. By linear analysis at those points, we show that these instantons connect the initial critical point of the dynamics, with at least one unstable direction, directly to the final fixed point. 
We also show that the memory content of these gates only affects the relaxation time to reach the logically consistent solution. Finally, we demonstrate, by solving the corresponding stochastic differential equations, that since instantons connect critical points, noise and perturbations may change the instanton trajectory in the phase space, but not the initial and final critical points. Therefore, even for extremely large noise levels, the gates self-organize to the correct solution.
Our work provides a physical understanding of, and can serve as an inspiration for, new models of bi-directional logic gates that are emerging as important tools in physics-inspired, unconventional computing. 
\end{abstract}    

                          
\maketitle

\section{Introduction}
Traditional Boolean logic is uni-directional, namely, given the truth value of a set of input terminals, one finds the consistent output value according to a given truth table~\cite{balabanian2007digital}. This is the type of logic that is employed, e.g., in our standard computing paradigm~\cite{parhami1999computer}. 

Recently, a new type of logic has been introduced by two of us (FL and MD)~\cite{dmm} that is both ``invertible'' and ``terminal-agnostic.'' This means that, in addition to working as traditional Boolean logic does from input terminals to output terminals, it can work ``in reverse,'' without reference to any particular set of terminals: by assigning a truth value to {\it any} terminal (even those at the traditional output), the gate is able to find a {\it logically consistent} truth assignment of the other terminals~\cite{dmm}. Of course, this logic is not necessarily bijective, because, in most cases, logic gates have a different number of terminals on one end of the gate than the other. 

The physical ingredient to realize such a framework is {\it time non-locality} (memory)~\cite{dmm}. Memory allows the system to {\it self-organize} into the correct truth value according to the initial conditions assigned~\cite{umm}. For this reason, these gates were named {\it self-organizing logic gates} (SOLGs)~\cite{dmm,manukian2017inversion}. 

Note that the self-organizing logic we consider here has no relation to the invertible universal Toffoli gate that is employed, e.g., in quantum computation \cite{toffoli}. Toffoli gates are truly one-to-one invertible, having 3-bit inputs and 3-bit outputs. On the other hand, SOLGs need only to satisfy the correct logical proposition, without a one-to-one relation between any number of input and output terminals. Instead, it is worth mentioning another type of bi-directional logic that has been recently discussed in Ref.~\cite{stochasticpbits} using stochastic units (called {\emph p}-bits). These units fluctuate among all possible consistent inputs. However, in contrast to that work, the invertible logic we consider here is {\it deterministic}.

\begin{figure}[t]
	\centering
	\vspace{-0.3cm}
	\includegraphics[width=6.75cm,trim={1.5cm 1.5cm 1.05cm 1.5cm},clip]{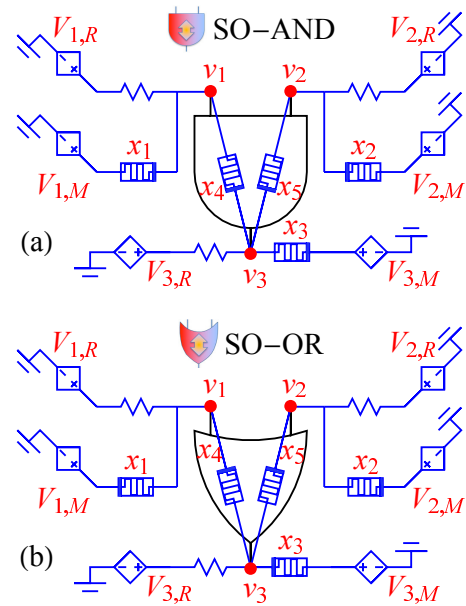}
	\vspace{-0.3cm}
	\caption{Circuit diagrams representing (a) a self-organizing AND (SO-AND) gate, and (b) a self-organizing OR (SO-OR) gate. The memristive elements (represented by a rectangle with a square waveform inside) have state variables $x_j$, $j=1,\dots,5$; an orientation, denoted by the bar on one side; and contain a parasitic capacitor in parallel. All resistors have the same resistance value. The form of the voltage-controlled voltage generators (represented by a diamond shape with $+$ and $-$ signs) is given in Table \ref{tab:coeff}. Voltages at the terminals are represented with a subscript $M$ if they are at a memristive terminal, with a subscript $R$ if they are at a resistance terminal.}
	\label{fig:circuits}
\end{figure}

With time being a fundamental ingredient, a dynamical systems approach is most natural to describe such gates. In particular, {\it non-linear} electronic (non-quantum) circuit elements with and without memory have been suggested as building blocks 
to realize SOLGs in practice~\cite{dmm} (see also Fig.~\ref{fig:circuits}). 

By assembling SOLGs with the appropriate architecture, one then obtains circuits that can solve complex problems efficiently by mapping the equilibrium (fixed) points of such circuits 
to the solution of the problem at hand, as shown in, e.g., Refs. \cite{dmm,manukian2017inversion, deeplearning,topo}. Moreover, it has been proved that, if those systems are engineered to be point dissipative \cite{hale_2010_asymptotic}, then, if equilibrium points are present, they do not show chaotic behavior \cite{no-chaos} or periodic orbits \cite{noperiod}. 

It was subsequently demonstrated~\cite{topo}, using topological field theory (TFT) applied to dynamical systems, that these circuits are described by a Witten-type TFT~\cite{Witten1}, and they support long-range order, mediated by {\it instantons}. Instantons are classical trajectories of the non-linear equations of motion describing these circuits (see, e.g.,~\cite{Coleman} or~\cite{Book1}). 

Instantons have been introduced first in the field of high-energy Physics to compute more efficiently tunneling matrix element 
between local vacua by a Wick rotation to Euclidean space~(see, e.g., \cite{InstQCD}). Local 
vacua are then transformed into critical points of the corresponding classical equations of motion~\cite{Coleman}. Therefore, instantons can be viewed as the 
classical analog of ``tunneling'' in the phase space. In fact, instantons can only connect critical points with different indices, namely different number 
of unstable directions. In turn, critical points can be located anywhere in the phase space. Therefore, instantons can be 
highly non-local objects. Finally, since critical points are related to the topology of the phase space, their number and index are robust against 
noise and perturbations~\cite{fomenko}. In other words, one needs to break the topology of the phase space to change its critical points. In practice, this 
requires changing the physical system itself. 

The intrinsic non-locality of instantons, coupled with the topological character of critical points, is reminiscent of the ``rigidity'' and topological 
character of the ground state of some strongly-correlated quantum systems that are currently investigated for topological quantum computation, namely 
quantum computation that is robust against dephasing and noise \cite{Freedman, Nayak, kitaev}. 
This analogy is not far-fetched. In fact, in the case of self-organizing circuits, instantons, by connecting topologically-distinct critical points in the phase space, correlate elements of the circuit non-locally in space and time \cite{topo}. This non-locality is somewhat reminiscent of quantum entanglement. However, SOLGs are circuits that achieve long-range order without quantum-mechanical effects.

The long-range order is not surprising since TFTs with condensed instantons are known to be log-conformal, hence support gapless excitations~\cite{Frankel}. Our previous work, however, leaves open the question as to whether the {\it single} SOLGs employ instantons as well, and, if so, what is the nature of the corresponding critical points. 

In this paper, we answer these questions by numerically solving the differential equations of self-organizing AND (SO-AND) and OR (SO-OR) gates {\it both with and without noise}. The set of Boolean operators $\{\text{AND, NOT}\}$ forms a functionally complete set, i.e., the two gates form a basis for all Boolean logic, as does the set $\{\text{OR, NOT}\}$. The NOT gate is implemented trivially in an electronic circuit, since it is simply a current (or voltage) inverter~\cite{neamen2001electronic}, and, therefore, it is not described herein. 

We find that the dynamics of these self-organizing gates proceeds as follows. Given an arbitrary initial condition the system ``scatters'' into unstable critical points whose unstable direction has an eigenvalue of the Jacobian matrix which is, in absolute value, considerably smaller than the largest eigenvalue of the stable directions. This makes them almost attractive to the initial dynamics. Subsequently, an instanton connects the unstable critical point to the equilibrium (fixed) point. In addition, the unstable direction evolves into 
a center manifold of the final fixed point. We also explicitly show, by perturbing the initial conditions and by solving stochastic differential equations, that although the trajectories connecting critical points may be substantially different due to either perturbations or noise, the instantons {\it always} go to the (unchanged) final fixed points. Using a simple diffusion model for typical memristors made of oxides, we relate the noise intensity to temperature. We find that even at very large temperatures (beyond the 
stability of the underlying materials) the SOLGs keep operating as expected. Therefore, the single logic units of more complicated self-organizing circuits take advantage of the instantonic long-range order, thus allowing the system to explore a vast phase space very efficiently, even in the presence of noise.

These results suggest that the topological features of SOLGs are essential in their operation as units of computation. In addition, our findings may  provide a physical understanding of other types of recently suggested (stochastic) bi-directional logic gates employed in 
unconventional computing~\cite{stochasticpbits,oscillatoryNetworks}.

\section{SOLGs formulation}
Let us start by outlining a model of SOLGs as a system of coupled, nonlinear, ordinary differential equations. We will then solve these equations numerically to identify instantons, their critical points, and, by diagonalizing the Jacobian (the matrix of the derivative of the flow vector field), their topological 
features. 

The implementation of SOLGs using electronic circuits is not unique, provided some of their properties are preserved~\cite{dmm,topo}. We refer the reader to Ref.~\cite{dmm} for all the mathematical properties of these gates. In order to make the phase space as small as possible -- hence the numerical analysis as easy as possible -- we choose a much simpler representation of SOLGs than that proposed in Ref. \cite{dmm}, which accomplishes the same tasks 
with a fewer number of variables~\footnote{Note, however, that in this very simplified representation of SOLGs, there may be stable critical points that do not satisfy Boolean logic. These cases are easily removed by adding voltage-controlled differential-current generators as in Ref.~\cite{dmm}. However, the subsequent increased dimensionality of the phase space would render the numerical analysis necessarily more complex.}.

\begin{table}[t]
	\centering
	\caption{Coefficients for the voltage-controlled voltage generators' relations given by $V_{i,j}=b_1 v_1+ b_2 v_2 + b_3 v_3 +dc_{gate}$, where 
		$i=1,2,3$ and $j=R,M$. All voltage are in Volts.}
	\label{tab:coeff}
	\begin{tabular}{|c|c|c|c|c|c|}
		\hline
		& $b_1$  & $b_2 $ &$b_3$  & $dc_{AND}$ & $dc_{OR}$\\ \hline
		$V_{1,M}$ & 0  & -1 & 1  & 1 & -1 \\ \hline
		$V_{1,R}$ & 3  & 1  & -2 & -1 & 1 \\ \hline
		$V_{2,M}$ & -1 & 0  & 1  & 1 & -1 \\ \hline
		$V_{2,R}$ & 1  & 3  & -2 & -1 & 1 \\ \hline
		$V_{3,M}$ & 2  & 2  & -1 & -2 & 2\\ \hline
		$V_{3,R }$& -3 & -3 & 5  & 2 &-2 \\ \hline
	\end{tabular}
\end{table}

In Fig. \ref{fig:circuits}, we show the SO-AND/OR gates we employ in this work. They are modeled with standard resistors, resistors with memory (memristive elements)~\cite{09_memelements}, and voltage-controlled voltage generators (VCVGs)~\cite{dmm}. The 
memristive elements contain a capacitance in parallel, representing parasitic capacitive effects. The difference between the circuitry of the SO-AND and SO-OR gates is the orientation of the memristive elements, and the definitions of the VCVGs (see Table \ref{tab:coeff}).

We want these gates to self-organize into the correct logical proposition irrespective of the terminal to which the truth value is assigned. To better 
understand how this is accomplished, it is beneficial to start from a specific example. Let us then choose to encode the logical 1 (True) with 1~V and the logical 0 (False) with $-1$~V. 

Consider first the SO-AND. 
If we set the voltage $v_1$ to 1 V, the system should evolve to either $v_2 = v_3= 1$ V or $v_2=v_3 = -1$ V. Both are logically consistent with an AND truth table. 
On the other hand, if we consider the SO-OR gate, and fix $v_1$ to $-1$~V (logical 0), the system should evolve to either  
$v_2=v_3=-1$~V or $v_2=v_3= 1$~V. The final result will depend on the initial conditions, 
namely the initial values of all voltages and internal state variables. 

Below, we describe a set of dynamical equations that accomplishes the above tasks. 
For the evolution of the memristive state variables we choose an equation of motion of the form~\cite{dmm},
\begin{equation}
\frac{d}{dt}x_j = -\alpha h(x_j, v_{M_j})g(x_j)v_{M_j},
\label{eq:memx}
\end{equation}
where $x_j$ is the state variable for the $j$-th memristive element. The function $h$ serves to cutoff the dynamics of the state variable in certain regimes. We choose the conductance of these elements, $g(x) = ((R_{off} - R_{on})x + R_{on})^{-1}$, where we set $R_{off}=1\ \Omega$ and $R_{on} =0.01\ \Omega$. Thus, $g(x)v_M$ is equal to the current flowing through a memristor. The voltage drop, $v_M$, is measured based on the orientation of the memristor: $v_M=v_a-v_b$, where $v_b$ is measured from the thick-bar side of the electronic symbol for the memristor. The coefficient $\alpha$ is restricted to be positive, and we choose $\alpha=60$. The physical meaning of $\alpha$ is discussed in Ref.~\cite{memelements}. Finally, the values of the state variables are bounded, and are typically chosen to be $x \in [0, 1]$~\cite{dmm}. 
\begin{figure}[t]
	\centering
	\includegraphics[width=3.5in]{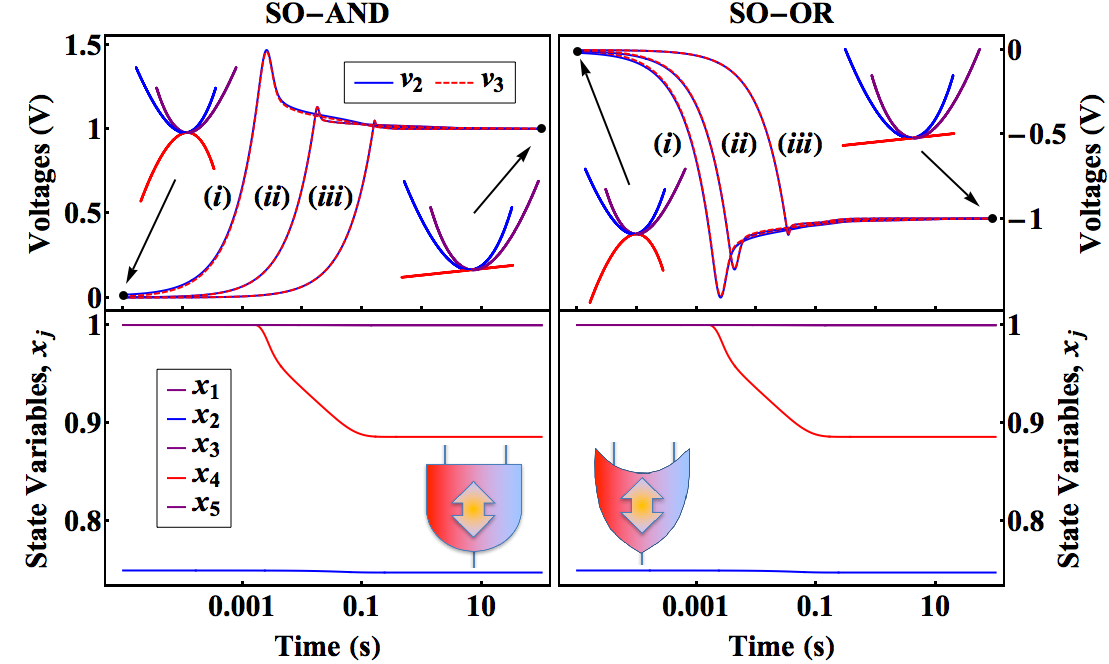}
	\vspace{-0.5cm}
	\caption{Time evolution of the voltages (top-left panel) and state variables (bottom-left panel) of the SO-AND compared to the time evolution of the voltages (top-right panel) and state variables (bottom-right panel) of the SO-OR. Elementary instantons in the SO-AND gate (top-left) are shown for ({\emph{i}}) $\delta v_2 = 10^{-2}V$, ({\emph{ii}}) $\delta v_2 =10^{-3}V$, ({\emph{iii}}) $\delta v_2 = 10^{-4}V$; Elementary instantons  in the SO-OR gate (top-right) are shown for ({\emph{i}}) $\delta v_2 = 10^{-2}V$, ({\emph{ii}}) $\delta v_2 =5\times10^{-3}V$, ({\emph{iii}}) $\delta v_2 = 5\times10^{-4}V$. For perturbation $\delta v_2 = 10^{-2}V$, the particular choice of initial conditions results in the memristor state variables evolving identically for both gates: (bottom-left) SO-AND; (bottom-right) SO-OR. Only memristors associated with $x_2$ and $x_4$ evolve in time. In addition, the voltage evolution of the SO-AND and the SO-OR are specularly symmetric (observe the  $\delta v_2 = 10^{-2}V$ case), as expected by their truth tables. The instanton connects the initial time critical point with two stable directions (positive curvature parabolas) and one unstable direction (negative curvature parabola), with the final state critical point with all three stable directions (fixed point). The unstable direction evolves into a center manifold (flat red line).}
	\label{fig:evo}
\end{figure}

Ideally, in order to strictly enforce $x \in [0, 1]$, $h(x, v_M)$ should be represented by step functions~\cite{dmm}. However, in practical realizations and numerical simulations, the step functions should be replaced by some differentiable 
function. We use,~\cite{dmm}
\begin{equation}
\begin{split}
h(x,v_M) =(1-e^{-kx})\hat{\theta}^r\left(\frac{v_M}{2V_t}\right) +\\ (1-e^{-k(1-x)})\hat{\theta}^r\left(-\frac{v_M}{2V_t}\right),
\end{split}
\label{eq:stepapprox}
\end{equation}
where $k=2$, and choose $V_t=0.1$ V. The $\hat{\theta}^r$ function is defined as,
\begin{equation}
\hat{\theta}^r(y) = \left\{
\begin{array}{ll}
1 & \quad y > 1 \\
\sum_{i=r+1}^{2r+1} a_i y^i & \quad 0 \leq y \leq 1\\
0 & \quad y < 0
\end{array}
\right.
\end{equation}
where we use the simplest case, $r=1$. The coefficients can be found by requiring continuity and differentiability in $y=0$ and $y=1$. This is equivalent to satisfying equations $\sum_{i=r+1}^{2r+1} a_i = 1$ and $\sum_{i=r+1}^{2r+1} {i \choose l} a_i = 0$ for $l=1,\ldots, r$. The coefficients for our implementation are $a_{2}=3$ and $a_{3}=-2$.
\begin{figure}[t]
	\centering
	\includegraphics[width=7.5cm]{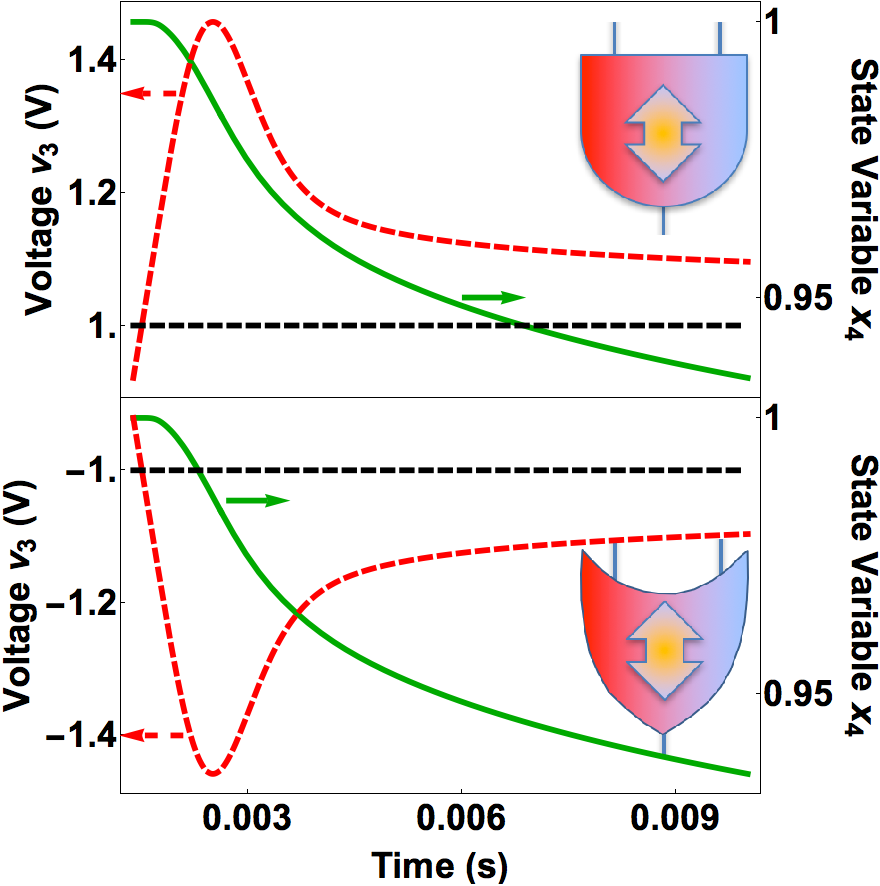}
	\vspace{-0.1cm}
	\caption{The elementary instanton in a single SOLG can best be understood from the restricted interval of the memristor state variable, $[0,1]$. The voltage shown is $v_3$, for a perturbation $\delta v_2=10^{-2}$ (see Fig.~\ref{fig:evo}). We see that for our particular initial conditions, the voltage must leave the $[-1\mathrm{V},1\mathrm{V}]$ interval for the memristor ($x_4$ is shown) to change its state.}
	\label{fig:inst}
\end{figure}

If we analyze the particular case discussed above, we fix, for both SOLGs, the voltage generator on terminal 1, and we perform standard nodal analysis on terminals 2 and 3 to find (see also Fig.~\ref{fig:circuits}),
\begin{equation}
\begin{split}
C(-\frac{d}{dt}v_ 1-2\frac{d}{dt}v_ 2+2\frac{d}{dt}{v}_3)=-i_{2}+\\
(v_ 2-v_3)g(x_5)+(-V_{2,M}+v_ 2)g(x_2)+\frac{-V_{2,R}+v_2}{R},\label{C1}
\end{split}
\end{equation}
\begin{equation}
\begin{split}
C(-3\frac{d}{dt}{v}_ 1-3\frac{d}{dt}{v}_ 2+4\frac{d}{dt}{v}_3)=-i_{3}+(v_ 1-v_3)g(x_4)+\\
\frac{V_{3,R}-v_3}{R}+(v_ 2-v_3)g(x_5)+(-v_3+V_{3,M})g(x_3),\label{C2}
\end{split}
\end{equation}
where the capacitance is $C=10^{-5}\ F$ and $R=1\ \Omega$ ~\footnote{In this example, we have used a large parasitic capacitance. However, this is not necessary as shown in Ref.~\cite{dmm}. A large capacitance value simply causes the peak of the instantonic trajectory to broaden, hence simplifying further the numerical analysis.}.
The VCVGs generate a voltage from the relation $
V_{i,j}=b_1 v_1+ b_2 v_2 + b_3 v_3 +dc_{gate}
$, 
with $dc_{gate}$ a constant voltage specific to each gate~\cite{dmm}. The coefficients, $b_k$, along with $dc_{gate}$, are given in Table~\ref{tab:coeff}. Terminals 2 and 3 are floating, therefore, $i_2=i_3=0$. Additionally, $\frac{d}{dt}v_1=0$, due to terminal 1 being attached to a voltage generator that is held constant. The role of the VCVGs 
is to inject a large current when the gate is in an inconsistent configuration, a small current otherwise. 

By solving numerically Eqs.~(\ref{eq:memx}), (\ref{C1}), and (\ref{C2}), with appropriate substitutions, we obtain precisely what we were after: a consistent logical solution for the given gate. This is reported in Fig.~\ref{fig:evo}, where, for the particular initial conditions chosen, 
we obtain a consistent solution for each SOLG: for the SO-AND, by starting at the logical 1 for $v_1$, we obtain the logical 1 at both 
$v_2$ and $v_3$. Instead, for the SO-OR, by starting at the logical 0 at $v_1$, we obtain the logical 0 at both $v_2$ and $v_{3}$. 

In the general case, the evolution of the terminal voltages and the memristive state variables of the SOLGs can be written compactly as, 
\begin{equation}
\dot{{\bf x}}(t) = {\bf F}({\bf x}(t)),\label{flow}
\end{equation}
where ${\bf x} = \{v_1, \ldots, v_m, x_1, \ldots, x_n\} \in X$ ($X$ is the phase space) represents the voltages, $v_j$, the internal state variables of the memristors, $x_j$, and ${\bf F}$ is a system of nonlinear ordinary differential equations, representing 
the flow vector field. 

The dynamical variables of the system then inhabit a phase space, $X\subset \mathbb{R}^{m+n}$. For the SO-AND/OR gates, $m=3$ and $n=5$. For the numerical simulations shown in Fig.~\ref{fig:evo} we have chosen to hold $v_1$ constant, 
so that the system has only seven dynamical variables.

\section{Instantons and stability analysis} Solutions ${\bf x}_{cr}$ to ${\bf F}({\bf x}_{cr}) = 0$ are the critical points in the phase space we are after. 
We have performed an extensive search of critical points of Eq.~\ref{flow} in the phase space, and found some with one unstable direction, and some with two unstable directions. Since our goal is simply to show that instantons are present in SOLGs, we focus on those originating from initial critical points with only one unstable direction. 

One such critical point is ${\bf x}_{cr}= \{v_2, v_{3}, x_1, x_2, x_3, x_4, x_5\} = \{0, 0, 1, 1, 0.75, 1, 1\}$. It is unstable if we hold $v_1=1$~V for the SO-AND, and  $v_1=-1$~V for the SO-OR. We check this by performing linear stability analysis, constructing the Jacobian matrix, $[J({\bf x})]_{ij} = \partial F_i({\bf x})/\partial x_j$, where differentiation is performed symbolically. We then determine, numerically, the eigenvalues of the Jacobian for the given critical point.
We then perturb the voltage on $v_2$ by, say, $\delta v_2=0.01$~V, causing the system to evolve via numerical integration to obtain the full dynamics shown in Fig.~\ref{fig:evo}. Results from various values of the perturbation $\delta v_2$ are also reported in Fig.~\ref{fig:evo}, explicitly showing the topological character of the solution search: the trajectories may depend strongly on perturbations, but not the critical points, hence the final solution.
 
The linearized equations around the critical points can be written as, $\dot{ {\bf x}} \approx {\bf J}({\bf x}_{cr})({\bf x} - {\bf x}_{cr})$, which result in the trajectories $ {\bf x}(t) \approx {\bf x}_{cr} + \sum_i {\bf v}_i e^{\lambda_i t}$. 
The sum is over eigenvalues $\lambda_i$ and associated eigenvectors ${\bf v}_i$. The eigenvectors corresponding to $\text{Re } \lambda_i < 0$ and $\text{Re }\lambda_i > 0$ define the vector spaces tangent to the stable and unstable manifolds, respectively, at each critical point. 

All eigenvectors with $\text{Re }\lambda_i= 0$ are associated to~\emph{center manifolds}. In our case these center manifolds arise from the indeterminacy of the internal state variables around a critical point. 
To illustrate this point better, consider the example shown in Fig.~\ref{fig:evo}, where we see that the system evolves between a critical point with a spectrum $\{\text{sign}(\lambda_i)\} =  \{-, -, +, 0, 0, 0, 0 \}$ to a final critical point $\{-, -, 0, 0, 0, 0, 0\}$, with all stable and center directions. The overall reduction of unstable directions is a general feature of the instantons. The resulting center manifolds do not change the stability profile of the critical point, but rather can be seen as the result of additional freedom the system has in order to satisfy the equilibrium condition ${\bf F (x)} = 0$. This freedom manifests itself in the morphing of the unstable direction of the initial point to 
a center manifold of the final equilibrium point. 

To better clarify how the dynamics of SOLGs result in the emergence of instantons, Fig. \ref{fig:inst} shows the time evolution of the memristor 
internal state variable $x_4$ (see also Fig.~\ref{fig:circuits}). This internal state does not evolve until $v_{3}$ exceeds the interval values 
$[-1\mathrm{V},1\mathrm{V}]$, allowing current to flow in the opposite direction through that memristive element. Only then can the memristor between terminals 1 and 3 of Fig.~\ref{fig:circuits} change its state, thus allowing a rapid variation of $v_3$ towards the equilibrium solution, hence the emergence of an instanton.
	 
\begin{figure}[t]
	\centering
	\includegraphics[width=7.5cm]{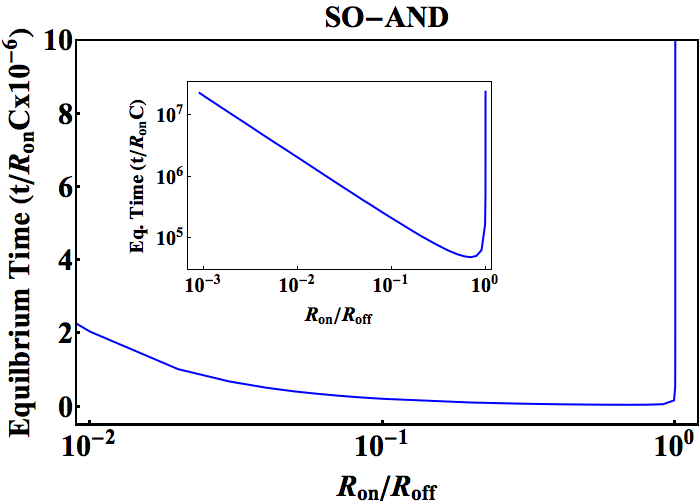}
	\vspace{-0.1cm}
	\caption{The memory content of the SOLGs ($R_{on}/R_{off}$) affects the time interval necessary to achieve equilibrium. The dependence of equilibrium time on memory content is illustrated in the main panel on a linear-log scale. Equilibrium is defined as the time necessary for voltages to be within 1\% of their steady-state values for a given ratio $R_{on}/R_{off}$ ($R_{on}=0.01 \ \Omega$ held fixed). The inset further illustrates the dependence on a log-log scale. The memory vanishes as $R_{on}/R_{off}\rightarrow1$, causing dynamics to slow to a halt, while the system approaches infinite memory as $R_{on}/R_{off}\rightarrow0$, causing the dynamics to slow as well.}
	\label{fig:ratio}
\end{figure}

Finally, to show the effect of memory on the dynamics of the SOLGs, we consider the ratio $R_{on}/R_{off}$ as a measure of memory. For $R_{on}/R_{off}\rightarrow1$, the system has vanishing memory; for  $R_{on}/R_{off}\rightarrow0$, the system  approaches infinite memory. In Fig. \ref{fig:ratio}, we see, as $R_{on}/R_{off}$ approaches either of the limiting values, the dynamics of the SOLGs slow down in reaching their logically-consistent (equilibrium) solution. (Equilibrium time is defined by the terminal voltages being within 1\% of steady-state values after the initial dynamics have settled.) In the limit of vanishing memory, the system lacks a mechanism to inject current, therefore, the SOLG loses the ability to self-organize. In the opposite limit of infinite memory, the dynamics are slowed due to the system possessing too many pathways to explore. The inset of Fig. \ref{fig:ratio} shows there is an optimum ratio for speeding up the dynamics, which will depend on the particular physical systems used to implement these gates.

Note that we have thus far assumed the system has found itself in an unstable critical point. An important question is how the SOLGs find their way to an unstable critical point from an arbitrary initial condition at $t=0$. The very presence of one or more unstable directions would make those critical points repulsive to the system, \emph{unless} the real part of the unstable eigenvalues were much smaller than the real part of the stable eigenvalues. This is indeed what 
we find in our simulations. For instance, for the critical point described above, the largest stable eigenvalue is of the order of $\sim 10^2$, and the magnitude of the unstable eigenvalue is $\sim 10^{-3}$. We find even larger orders of magnitude differences for the other critical points we have analyzed. This makes these critical points \emph{almost} attractive, or 
at least not repulsive enough to prevent the system from falling into them.

\section{SOLGs with noise} As further evidence of topological robustness, we test the performance of the SOLGs under the influence of additive noise, modeling the internal noise of the memristors. The compact representation of the system is reformulated as,
\begin{equation}
	\dot{{\bf x}}(t) = {\bf F}({\bf x}(t))+{\bf \xi}(t),
	\label{flownoise}
\end{equation}
where ${\bf \xi}=\{0,0,\xi_1,\xi_2,\xi_3,\xi_4,\xi_5\}$. That is, the additive noise appears only in the equations for the memristor state variables,
\begin{equation}
\frac{d}{dt}x_j = -\alpha h(x_j, v_{M_j})g(x_j)v_{M_j}+\xi_j(t),
\label{eq:memxnoise}
\end{equation}
where $\xi_j(t)$ is a white noise process of intensity $\Gamma$, with properties,
\begin{equation}
\langle{\bf \xi}_j(t)\rangle=0, \;\;\;
\langle{\bf \xi}_i(t){\bf \xi}_j(t')\rangle=\Gamma\delta(t-t')\delta_{i,j}.
\label{mean}
\end{equation}

\begin{figure}[b]
	\centering
	\includegraphics[width=7.5cm]{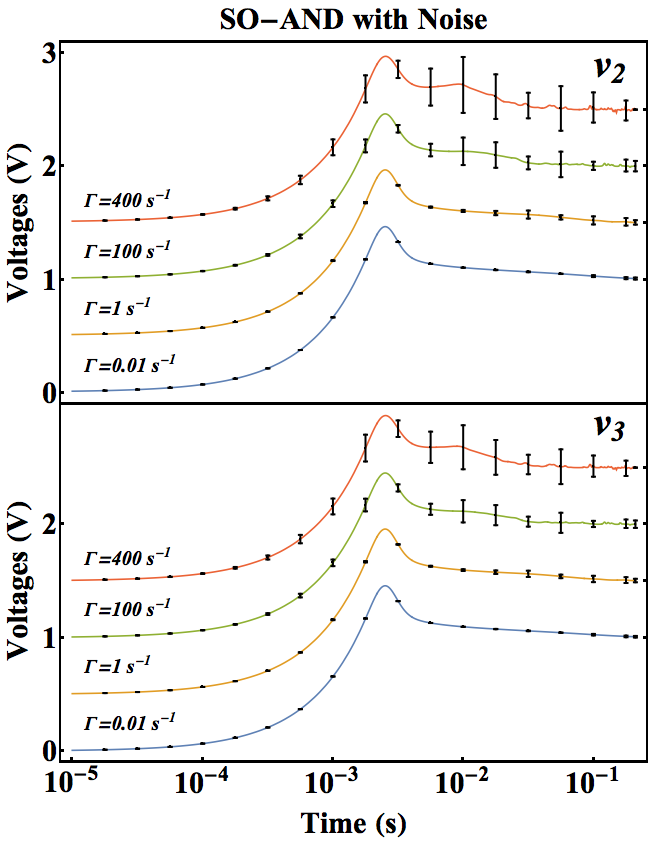}
	\vspace{-0.1cm}
	\caption{SOLGs are topologically robust in the presence of noise, meaning that the critical points connected by the instanton are unchanged, though the trajectory is modified. Using the same initial conditions as used in curve (\emph{i}) in the top-left panel of Fig. \ref{fig:evo}, we have added white noise of varying strength to the simulation of the SO-AND gate's memristors. The terminal voltages are shown: $v_2$ (top panel) and $v_3$ (bottom panel). Each curve is the average of 100 simulations, with curves being translated upward for the purpose of clarity. Error bars have a height of 2 standard deviations, with some so small that they appear to be horizontal black lines. The ticks on the right axes mark the location of 1 V with respect to each corresponding curve. Note that dynamics are shown up to the equilibrium time associated with $R_{on}/R_{off}=0.01$ in Fig. \ref{fig:ratio}.} 
	\label{fig:noise}
\end{figure}

 Simulations of the SO-AND gate with noise are shown in Fig. \ref{fig:noise}. Each curve is the average of 100 simulations, with  each simulation having the same parameters and initial conditions as curve (\emph{i}) in the top-left panel of Fig. \ref{fig:evo}. Notice that the error bars are largest after the voltage peak in the dynamics, then the error overall decreases as time increases. This is consistent with the larger instability of the trajectory when it changes more rapidly. 
 
 Increasing the noise intensity beyond the $\Gamma=400$ s$^{-1}$ value results in dynamics that are no longer associated with the phase space specified above. 
 This can be understood by recalling that the function $h(x,v_M)$ in Eq.~(\ref{eq:memx}) cuts off the memristor dynamics to enforce $x\in[0,1]$. By increasing the 
 noise level beyond $\Gamma=400$ s$^{-1}$ (with $R_{on}/R_{off}=10^{-2}$), the internal states are driven far beyond the physical limit of $x\in[0,1]$. For example, 
 in a TiO$_2$ memristor~\cite{missingmem}, the state variable is a measure of oxygen vacancies in the semiconductor film, so the boundaries of the state variable are well-defined. A noise of such an intensity as to move the state variables beyond their bounds, would imply a physical destruction of the device. 
However, it is clear from Fig.~\ref{fig:noise} that even at a high level of intensity, the noise has not destroyed the critical points, confirming that in order to change the number and character of the critical points, the topology has to change drastically.

To make contact with actual experiments we estimate the temperature as a function of the noise strength. We, again, refer to TiO$_2$ memristors~\cite{missingmem}. In that case, the intensity of the noise is related to the diffusion coefficient,
 \begin{equation}
D=\frac{\Gamma L^2}{2},
 \label{diff}
 \end{equation}
  where $L$ is the length of the oxide region of the memristor. Additionally, the diffusion coefficient is related to temperature $T$,
  \begin{equation}
 D=D_0 exp\left(-\frac{E_{\nu}}{k_BT}\right),
 \label{temp}
 \end{equation}
 where $D_0=10^{-3}$ cm$^2$/s is the maximal diffusion coefficient, $E_{\nu}=0.5$ eV is the activation energy for oxygen vacancy diffusion, and we take $L=100$ nm  \cite{stochasticmemory,tio2}. 
 Therefore, we can associate a temperature with the noise strength $\Gamma$. We find $T= 271$ K for $\Gamma=0.01$ s$^{-1}$; $T= 345$ K for $\Gamma=1$ s$^{-1}$; $T= 475$ K for $\Gamma=100$ s$^{-1}$; $T= 536$ K for $\Gamma=400$ s$^{-1}$. These temperature estimates indicate that 
 the largest value of the noise intensity is likely within the parameters of physical instability of these types of memristors.

\

\section{Conclusions} In this work we have shown that the recently suggested self-organizing logic gates (which, unlike standard uni-directional Boolean gates, are ``terminal-agnostic'') use instantons to slice through the 
(large) phase space to find the stable equilibria corresponding to the consistent logical solutions of the Boolean gate they represent. 
The elementary instantons that are generated during the dynamics of these gates directly connect  unstable initial-state critical points with the stable equilibrium points, and eliminate the unstable directions by morphing them into center manifolds. The stable equilibria are the result of the parameter freedom of the internal state variables. 

We have also demonstrated explicitly that the memory content of these gates only changes the time scale to reach the logically-consistent equilibria, while perturbations and noise can only change the trajectories in phase space but not the initial and final critical points. This implies again the topological robustness of these SOLGs.

This work then provides a better understanding of self-organizing logic, and may prove useful in the design of other practical realizations of this framework. Additionally, this work could explain other types of bi-directional logic that are being developed in the context of unconventional computing. 

\begin{acknowledgments} We thank Forrest Sheldon for useful discussions. This material is based upon work supported by the National Science Foundation Graduate Research Fellowship under Grant No. DGE-1650112. S.B. acknowledges support from the Alfred P. Sloan Foundation's Minority Ph.D. Program. H.M. acknowledges support from a DoD-SMART fellowship.
\end{acknowledgments}


\begin{thebibliography}{19}%
	\makeatletter
	\providecommand \@ifxundefined [1]{%
		\@ifx{#1\undefined}
	}%
	\providecommand \@ifnum [1]{%
		\ifnum #1\expandafter \@firstoftwo
		\else \expandafter \@secondoftwo
		\fi
	}%
	\providecommand \@ifx [1]{%
		\ifx #1\expandafter \@firstoftwo
		\else \expandafter \@secondoftwo
		\fi
	}%
	\providecommand \natexlab [1]{#1}%
	\providecommand \enquote  [1]{``#1''}%
	\providecommand \bibnamefont  [1]{#1}%
	\providecommand \bibfnamefont [1]{#1}%
	\providecommand \citenamefont [1]{#1}%
	\providecommand \href@noop [0]{\@secondoftwo}%
	\providecommand \href [0]{\begingroup \@sanitize@url \@href}%
	\providecommand \@href[1]{\@@startlink{#1}\@@href}%
	\providecommand \@@href[1]{\endgroup#1\@@endlink}%
	\providecommand \@sanitize@url [0]{\catcode `\\12\catcode `\$12\catcode
		`\&12\catcode `\#12\catcode `\^12\catcode `\_12\catcode `\%12\relax}%
	\providecommand \@@startlink[1]{}%
	\providecommand \@@endlink[0]{}%
	\providecommand \url  [0]{\begingroup\@sanitize@url \@url }%
	\providecommand \@url [1]{\endgroup\@href {#1}{\urlprefix }}%
	\providecommand \urlprefix  [0]{URL }%
	\providecommand \Eprint [0]{\href }%
	\providecommand \doibase [0]{http://dx.doi.org/}%
	\providecommand \selectlanguage [0]{\@gobble}%
	\providecommand \bibinfo  [0]{\@secondoftwo}%
	\providecommand \bibfield  [0]{\@secondoftwo}%
	\providecommand \translation [1]{[#1]}%
	\providecommand \BibitemOpen [0]{}%
	\providecommand \bibitemStop [0]{}%
	\providecommand \bibitemNoStop [0]{.\EOS\space}%
	\providecommand \EOS [0]{\spacefactor3000\relax}%
	\providecommand \BibitemShut  [1]{\csname bibitem#1\endcsname}%
	\let\auto@bib@innerbib\@empty
	\bibitem [{\citenamefont {Balabanian}\ and\ \citenamefont
		{Carlson}(2007)}]{balabanian2007digital}%
	\BibitemOpen
	\bibfield  {author} {\bibinfo {author} {\bibfnamefont {N.}~\bibnamefont
			{Balabanian}}\ and\ \bibinfo {author} {\bibfnamefont {B.}~\bibnamefont
			{Carlson}},\ }\href@noop {} {\emph {\bibinfo {title} {Digital Logic Design
				Principles}}}\ (\bibinfo  {publisher} {John Wiley \& Sons},\ \bibinfo {year}
	{2007})\BibitemShut {NoStop}%
	\bibitem [{\citenamefont {Parhami}()}]{parhami1999computer}%
	\BibitemOpen
	\bibfield  {author} {\bibinfo {author} {\bibfnamefont {B.}~\bibnamefont
			{Parhami}},\ }\href@noop {} {\emph {\bibinfo {title} {{Computer
					arithmetic}}}},\ Vol.~\bibinfo {volume} {20}\BibitemShut {NoStop}%
	\bibitem [{\citenamefont {Traversa}\ and\ \citenamefont {{Di
				Ventra}}(2017)}]{dmm}%
	\BibitemOpen
	\bibfield  {author} {\bibinfo {author} {\bibfnamefont {F.~L.}\ \bibnamefont
			{Traversa}}\ and\ \bibinfo {author} {\bibfnamefont {M.}~\bibnamefont {{Di
					Ventra}}},\ }\bibfield  {title} {\enquote {\bibinfo {title} {Polynomial-time
				solution of prime factorization and np-complete problems with digital
				memcomputing machines},}\ }\href {\doibase 10.1063/1.4975761} {\bibfield
		{journal} {\bibinfo  {journal} {Chaos: An Interdisciplinary Journal of
				Nonlinear Science}\ }\textbf {\bibinfo {volume} {27}},\ \bibinfo {pages}
		{023107} (\bibinfo {year} {2017})},\ \Eprint
	{http://arxiv.org/abs/http://dx.doi.org/10.1063/1.4975761}
	{http://dx.doi.org/10.1063/1.4975761} \BibitemShut {NoStop}%
	\bibitem [{\citenamefont {Traversa}\ and\ \citenamefont {{Di
				Ventra}}(2015)}]{umm}%
	\BibitemOpen
	\bibfield  {author} {\bibinfo {author} {\bibfnamefont {F.~L.}\ \bibnamefont
			{Traversa}}\ and\ \bibinfo {author} {\bibfnamefont {M.}~\bibnamefont {{Di
					Ventra}}},\ }\bibfield  {title} {\enquote {\bibinfo {title} {Universal
				memcomputing machines},}\ }\href@noop {} {\bibfield  {journal} {\bibinfo
			{journal} {IEEE Trans. on Neural Networks}\ }\textbf {\bibinfo {volume}
			{26}},\ \bibinfo {pages} {2702--2715} (\bibinfo {year} {2015})}\BibitemShut
	{NoStop}%
	\bibitem [{\citenamefont {Manukian}\ \emph {et~al.}(2017)\citenamefont
		{Manukian}, \citenamefont {Traversa},\ and\ \citenamefont {{Di
				Ventra}}}]{manukian2017inversion}%
	\BibitemOpen
	\bibfield  {author} {\bibinfo {author} {\bibfnamefont {H.}~\bibnamefont
			{Manukian}}, \bibinfo {author} {\bibfnamefont {F.~L.}\ \bibnamefont
			{Traversa}}, \ and\ \bibinfo {author} {\bibfnamefont {M.}~\bibnamefont {{Di
					Ventra}}},\ }\bibfield  {title} {\enquote {\bibinfo {title} {Memcomputing
				numerical inversion with self-organizing logic gates},}\ }\href {\doibase
		10.1109/TNNLS.2017.2697386} {\bibfield  {journal} {\bibinfo  {journal} {IEEE
				Transactions on Neural Networks and Learning Systems}\ }\textbf {\bibinfo
			{volume} {PP}},\ \bibinfo {pages} {1--6} (\bibinfo {year}
		{2017})}\BibitemShut {NoStop}%
	\bibitem [{\citenamefont {Toffoli}\ (1980)\citenamefont
		{Toffoli}}]{toffoli}%
	\BibitemOpen
	\bibfield  {author} {\bibinfo {author} {\bibfnamefont {T.}~\bibnamefont
			{Toffoli}},\ }\bibfield  {title}
	{\enquote {\bibinfo {title} {Resersible Computing},}\ }in\ \href {\doibase
		10.1007/3-540-10003-2\_104} {\emph {\bibinfo {booktitle} {International Colloquium on Automata, Languages, and Programming (ICALP)}}}\ (\bibinfo {year} {1980})\
	pp.\ \bibinfo {pages} {632--644}\BibitemShut {NoStop}%
	\bibitem [{\citenamefont {Camsari}\ \emph {et~al.}(2017)\citenamefont
		{Camsari}, \citenamefont {Faria}, \citenamefont {Sutton},\ and\ \citenamefont
		{Datta}}]{stochasticpbits}%
	\BibitemOpen
	\bibfield  {author} {\bibinfo {author} {\bibfnamefont {K.~Y.}\ \bibnamefont
			{Camsari}}, \bibinfo {author} {\bibfnamefont {R.}~\bibnamefont {Faria}},
		\bibinfo {author} {\bibfnamefont {B.~M.}\ \bibnamefont {Sutton}}, \ and\
		\bibinfo {author} {\bibfnamefont {S.}~\bibnamefont {Datta}},\ }\bibfield
	{title} {\enquote {\bibinfo {title} {Stochastic $p$-bits for invertible
				logic},}\ }\href {\doibase 10.1103/PhysRevX.7.031014} {\bibfield  {journal}
		{\bibinfo  {journal} {Phys. Rev. X}\ }\textbf {\bibinfo {volume} {7}},\
		\bibinfo {pages} {031014} (\bibinfo {year} {2017})}\BibitemShut {NoStop}%
	\bibitem [{\citenamefont {Manukian}\ \emph {et~al.}(2018)\citenamefont
		{Manukian}, \citenamefont {Traversa},\ and\ \citenamefont
		{Di Ventra}}]{deeplearning}%
	\BibitemOpen
	\bibfield  {author} {\bibinfo {author} {\bibfnamefont {H.}\ \bibnamefont
			{Manukian}}, \bibinfo {author} {\bibfnamefont {F.~L.}~\bibnamefont {Traversa}},\ and\
		\bibinfo {author} {\bibfnamefont {M.}~\bibnamefont {Di Ventra}},\ }\bibfield
	{title} {\enquote {\bibinfo {title} {Accelerating Deep Learning with Memcomputing},}\ } {\bibfield  {journal}
		{\bibinfo  {journal} {e-print},\ }
		\bibinfo {pages} {arxiv:1801.00512}}\BibitemShut {NoStop}%
	\bibitem [{\citenamefont {{Di Ventra}}\ \emph {et~al.}(2017)\citenamefont {{Di
				Ventra}}, \citenamefont {Traversa},\ and\ \citenamefont
		{Ovchinnikov}}]{topo}%
	\BibitemOpen
	\bibfield  {author} {\bibinfo {author} {\bibfnamefont {M.}~\bibnamefont {{Di
					Ventra}}}, \bibinfo {author} {\bibfnamefont {F.~L.}\ \bibnamefont
			{Traversa}}, \ and\ \bibinfo {author} {\bibfnamefont {I.~V.}\ \bibnamefont
			{Ovchinnikov}},\ }\bibfield  {title} {\enquote {\bibinfo {title} {Topological
				field theory and computing with instantons},}\ }\href {\doibase
		10.1002/andp.201700123} {\bibfield  {journal} {\bibinfo  {journal} {Annalen
				der Physik}\ ,\ \bibinfo {pages} {1700123}} (\bibinfo {year}
		{2017})}\BibitemShut {NoStop}%
	\bibitem [{\citenamefont {Hale}(2010)}]{hale_2010_asymptotic}%
	\BibitemOpen
	\bibfield  {author} {\bibinfo {author} {\bibfnamefont {J.K.}\ \bibnamefont
			{Hale}},\ }\href@noop {} {\emph {\bibinfo {title} {Asymptotic Behavior of
				Dissipative Systems}}},\ \bibinfo {edition} {2nd}\ ed.,\ \bibinfo {series}
	{Mathematical Surveys and Monographs}, Vol.~\bibinfo {volume} {25}\ (\bibinfo
	{publisher} {American Mathematical Society},\ \bibinfo {address}
	{Providence, Rhode Island},\ \bibinfo {year} {2010})\BibitemShut {NoStop}%
\bibitem{no-chaos}
\bibinfo{author}{M. Di~Ventra} and \bibinfo{author}{F.L. Traversa,}
\newblock \bibinfo{title}{``Absence of chaos in digital memcomputing machines
	with solutions''},
\newblock \bibinfo{journal}{Phys. Lett. A}
\textbf{\bibinfo{volume}{381}}, \bibinfo{pages}{3255} (\bibinfo{year}{2017}).
\bibitem{noperiod}
\bibinfo{author}{M. {Di Ventra}} and \bibinfo{author}{F.L. Traversa,}
\newblock \bibinfo{title}{``Absence of periodic orbits in digital memcomputing
	machines with solutions''},
\newblock \bibinfo{journal}{Chaos: An Interdisciplinary Journal of
		Nonlinear Science} \textbf{\bibinfo{volume}{27}}, \bibinfo{pages}{101101}
(\bibinfo{year}{2017}).		
	\bibitem [{\citenamefont {Witten}(1988)}]{Witten1}%
	\BibitemOpen
	\bibfield  {author} {\bibinfo {author} {\bibfnamefont {E.}~\bibnamefont
			{Witten}},\ }\bibfield  {title} {\enquote {\bibinfo {title} {Topological
				quantum field theory},}\ }\href@noop {} {\bibfield  {journal} {\bibinfo
			{journal} {Comms. in Math. Phys.}\ }\textbf {\bibinfo {volume} {117}},\
		\bibinfo {pages} {353–386} (\bibinfo {year} {1988})}\BibitemShut {NoStop}%
	\bibitem [{\citenamefont {Coleman}(1977)}]{Coleman}%
	\BibitemOpen
	\bibfield  {author} {\bibinfo {author} {\bibfnamefont {S.}~\bibnamefont
			{Coleman}},\ }\href@noop {} {\emph {\bibinfo {title} {Aspects of Symmetry,
				Chapter 7}}}\ (\bibinfo  {publisher} {Cambridge University Press},\ \bibinfo
	{year} {1977})\BibitemShut {NoStop}%
	\bibitem [{\citenamefont {Hori}\ \emph {et~al.}(2000)\citenamefont {Hori},
		\citenamefont {Katz}, \citenamefont {Klemm}, \citenamefont {Thomas},
		\citenamefont {Vafa}, \citenamefont {Vakil},\ and\ \citenamefont
		{Zaslow}}]{Book1}%
	\BibitemOpen
	\bibfield  {author} {\bibinfo {author} {\bibfnamefont {K.}~\bibnamefont
			{Hori}}, \bibinfo {author} {\bibfnamefont {S.}~\bibnamefont {Katz}}, \bibinfo
		{author} {\bibfnamefont {R.}~\bibnamefont {Klemm}, \bibfnamefont
			{A.~Pandharipande}}, \bibinfo {author} {\bibfnamefont {R.}~\bibnamefont
			{Thomas}}, \bibinfo {author} {\bibfnamefont {C.}~\bibnamefont {Vafa}},
		\bibinfo {author} {\bibfnamefont {R.}~\bibnamefont {Vakil}}, \ and\ \bibinfo
		{author} {\bibfnamefont {E.}~\bibnamefont {Zaslow}},\ }\href@noop {} {\emph
		{\bibinfo {title} {Mirror symmetry}}}\ (\bibinfo  {publisher} {Clay
		Mathematics},\ \bibinfo {year} {2000})\BibitemShut {NoStop}%
	\bibitem [{\citenamefont {Shafer}\ \emph {et~al.}(1998)\citenamefont
		{Shafer},\ and\ \citenamefont
		{Shuryak}}]{InstQCD}%
	\BibitemOpen
	\bibfield  {author} {\bibinfo {author} {\bibfnamefont {T.}\ \bibnamefont
			{Schafer}}, \ and\
		\bibinfo {author} {\bibfnamefont {E.~V.}~\bibnamefont {Shuryak}},\ }\bibfield
	{title} {\enquote {\bibinfo {title} {Instantons in QCD},}\ }\href {\doibase 10.1103/RevModPhys.70.323} {\bibfield  {journal}
		{\bibinfo  {journal} {Rev. Mod. Phys.}\ }\textbf {\bibinfo {volume} {70}},\
		\bibinfo {pages} {323} (\bibinfo {year} {1998})}\BibitemShut {NoStop}%
	\bibitem [{\citenamefont {Fomenko}(1994)}]{fomenko}%
	\BibitemOpen
	\bibfield  {author} {\bibinfo {author} {\bibfnamefont {A.~T.}~\bibnamefont
			{Fomenko}},\ }\href@noop {} {\emph {\bibinfo {title} {Visual Geometry and Topology}}}\ (\bibinfo  {publisher} {Springer-Verlag Berlin Heidelberg},\ \bibinfo {year}
	{1994})\BibitemShut {NoStop}%
	\bibitem [{\citenamefont {Freedman}\ \emph {et~al.}(2002)\citenamefont
		{Freedman}, \citenamefont {Kitaev}, \citenamefont {Larsen},\ and\ \citenamefont {Wang}}]{Freedman}%
	\BibitemOpen
	\bibfield  {author} {\bibinfo {author} {\bibfnamefont {M.~H.}~\bibnamefont
			{Freedman}}, \bibinfo {author} {\bibfnamefont {A.}~\bibnamefont {Kitaev}}, \bibinfo {author} {\bibfnamefont {M.~J.}~\bibnamefont {Larsen}}, \
		and\ \bibinfo {author} {\bibfnamefont {Z.}~\bibnamefont {Wang}},\
	}\bibfield  {title} {\enquote {\bibinfo {title} {Topological quantum computing},}\ }\href {\doibase 10.1090/S0273-0979-02-00964-3 } {\bibfield  {journal}
		{\bibinfo  {journal} {Bullet. of Am. Math. Soc.}\ }\textbf {\bibinfo {volume} {40}},\
		\bibinfo {pages} {31-38} (\bibinfo {year} {2002})}\BibitemShut {NoStop}%
	\bibitem [{\citenamefont {Nayak}\ \emph {et~al.}(2008)\citenamefont
		{Nayak}, \citenamefont {Simon}, \citenamefont {Stern}, \citenamefont {Freedman},\ and\ \citenamefont {Das Sarma}}]{Nayak}%
	\BibitemOpen
	\bibfield  {author} {\bibinfo {author} {\bibfnamefont {C.}~\bibnamefont
			{Nayak}}, \bibinfo {author} {\bibfnamefont {S.~H.}~\bibnamefont {Simon}}, \bibinfo {author} {\bibfnamefont {A.}~\bibnamefont {Stern}}, \bibinfo {author} {\bibfnamefont {M.}~\bibnamefont {Freedman}}, \
		and\ \bibinfo {author} {\bibfnamefont {S.}~\bibnamefont {Das Sarma}},\
	}\bibfield  {title} {\enquote {\bibinfo {title} {Non-Abelian anyons and topological quantum computation},}\ }\href {\doibase 10.1103/RevModPhys.80.1083} {\bibfield  {journal}
		{\bibinfo  {journal} {Rev. Mod. Phys.}\ }\textbf {\bibinfo {volume} {80}},\
		\bibinfo {pages} {1083} (\bibinfo {year} {2008})}\BibitemShut {NoStop}%
	\bibitem [{\citenamefont {Kitaev}\ (2003)\citenamefont
		{Kitaev}}]{kitaev}%
	\BibitemOpen
	\bibfield  {author} {\bibinfo {author} {\bibfnamefont {A.}~\bibnamefont
			{Kitaev}},\
	}\bibfield  {title} {\enquote {\bibinfo {title} {Fault-tolerant quantum computation by anyons},}\ }\href {\doibase 10.1016/S0003-4916(02)00018-0} {\bibfield  {journal}
		{\bibinfo  {journal} {Annals of Physics}\ }\textbf {\bibinfo {volume} {303}},\
		\bibinfo {pages} {2-30} (\bibinfo {year} {2003})}\BibitemShut {NoStop}%
	\bibitem [{\citenamefont {Frenkel}\ \emph {et~al.}(2007)\citenamefont
		{Frenkel}, \citenamefont {Losev},\ and\ \citenamefont {Nekrasov}}]{Frankel}%
	\BibitemOpen
	\bibfield  {author} {\bibinfo {author} {\bibfnamefont {E.}~\bibnamefont
			{Frenkel}}, \bibinfo {author} {\bibfnamefont {A.}~\bibnamefont {Losev}}, \
		and\ \bibinfo {author} {\bibfnamefont {N.}~\bibnamefont {Nekrasov}},\
	}\bibfield  {title} {\enquote {\bibinfo {title} {Notes on instantons in
				topological field theory and beyond},}\ }\href@noop {} {\bibfield  {journal}
		{\bibinfo  {journal} {Nucl.~Phys.~B}\ }\textbf {\bibinfo {volume} {171}},\
		\bibinfo {pages} {215} (\bibinfo {year} {2007})}\BibitemShut {NoStop}%
	\bibitem [{\citenamefont {Neamen}()}]{neamen2001electronic}%
	\BibitemOpen
	\bibfield  {author} {\bibinfo {author} {\bibfnamefont {D.~A.}\ \bibnamefont
			{Neamen}},\ }\href@noop {} {\emph {\bibinfo {title} {Electronic circuit
				analysis and design}}},\ Vol.~\bibinfo {volume} {2}\ (\bibinfo  {publisher}
	{McGraw-Hill})\BibitemShut {NoStop}%
	\bibitem [{\citenamefont {Parihar}\ \emph {et~al.}(2017)\citenamefont
		{Parihar}, \citenamefont {Shukla}, \citenamefont {Jerry}, \citenamefont
		{Datta},\ and\ \citenamefont {Raychowdhury}}]{oscillatoryNetworks}%
	\BibitemOpen
	\bibfield  {author} {\bibinfo {author} {\bibfnamefont {A.}~\bibnamefont
			{Parihar}}, \bibinfo {author} {\bibfnamefont {N.}~\bibnamefont {Shukla}},
		\bibinfo {author} {\bibfnamefont {M.}~\bibnamefont {Jerry}}, \bibinfo
		{author} {\bibfnamefont {S.}~\bibnamefont {Datta}}, \ and\ \bibinfo {author}
		{\bibfnamefont {A.}~\bibnamefont {Raychowdhury}},\ }\bibfield  {title}
	{\enquote {\bibinfo {title} {Computational paradigms using oscillatory
				networks based on state-transition devices},}\ }in\ \href {\doibase
		10.1109/IJCNN.2017.7966285} {\emph {\bibinfo {booktitle} {2017 International
				Joint Conference on Neural Networks (IJCNN)}}}\ (\bibinfo {year} {2017})\
	pp.\ \bibinfo {pages} {3415--3422}\BibitemShut {NoStop}%
	\bibitem [{Note1()}]{Note1}%
	\BibitemOpen
	\bibinfo {note} {Note, however, that in this very simplified representation
		of SOLGs, there may be stable critical points that do not satisfy Boolean
		logic. These cases are easily removed by adding voltage-controlled
		differential-current generators as in Ref.~\cite {dmm}. However, the
		subsequent increased dimensionality of the phase space would render the
		numerical analysis necessarily more complex.}\BibitemShut {Stop}%
	\bibitem [{\citenamefont {{Di Ventra}}\ \emph {et~al.}(2009)\citenamefont {{Di
				Ventra}}, \citenamefont {Pershin},\ and\ \citenamefont
		{Chua}}]{09_memelements}%
	\BibitemOpen
	\bibfield  {author} {\bibinfo {author} {\bibfnamefont {M.}~\bibnamefont {{Di
					Ventra}}}, \bibinfo {author} {\bibfnamefont {Y.V.}\ \bibnamefont {Pershin}},
		\ and\ \bibinfo {author} {\bibfnamefont {L.O.}\ \bibnamefont {Chua}},\
	}\bibfield  {title} {\enquote {\bibinfo {title} {{Circuit Elements With
					Memory: Memristors, Memcapacitors, and Meminductors}},}\ }\href {\doibase
		10.1109/JPROC.2009.2021077} {\bibfield  {journal} {\bibinfo  {journal}
			{Proceedings of the IEEE}\ }\textbf {\bibinfo {volume} {97}},\ \bibinfo
		{pages} {1717--1724} (\bibinfo {year} {2009})}\BibitemShut {NoStop}%
	\bibitem [{\citenamefont {{Di Ventra}}\ and\ \citenamefont
		{Pershin}(2013)}]{memelements}%
	\BibitemOpen
	\bibfield  {author} {\bibinfo {author} {\bibfnamefont {M.}~\bibnamefont {{Di
					Ventra}}}\ and\ \bibinfo {author} {\bibfnamefont {Y.~V.}\ \bibnamefont
			{Pershin}},\ }\bibfield  {title} {\enquote {\bibinfo {title} {On the physical
				properties of memristive, memcapacitive and meminductive systems},}\
	}\href@noop {} {\bibfield  {journal} {\bibinfo  {journal} {Nanotechnology}\
		}\textbf {\bibinfo {volume} {24}} (\bibinfo {year} {2013})}\BibitemShut
	{NoStop}%
	\bibitem [{Note2()}]{Note2}%
	\BibitemOpen
	\bibinfo {note} {In this example, we have used a large parasitic capacitance.
		However, this is not necessary as shown in Ref.~\cite {dmm}. A large
		capacitance value simply causes the peak of the instantonic trajectory to
		broaden, hence simplifying further the numerical analysis.}\BibitemShut
	{NoStop}%
	\bibitem [{\citenamefont {Strukov}\ \emph {et~al.}(2008)\citenamefont
		{Strukov}, \citenamefont {Snider}, \citenamefont {Stewart},\ and\ \citenamefont
		{Williams}}]{missingmem}%
	\BibitemOpen
	\bibfield  {author} {\bibinfo {author} {\bibfnamefont {D.~B.}\ \bibnamefont
			{Strukov}}, \bibinfo {author} {\bibfnamefont {G.~S.}~\bibnamefont {Snider}},  \bibinfo {author} {\bibfnamefont {D.~R.}~\bibnamefont {Stewart}}, \ and\
		\bibinfo {author} {\bibfnamefont {R.~S.}~\bibnamefont {Williams}},\ }\bibfield
	{title} {\enquote {\bibinfo {title} {The missing memristor found},}\ }\href {\doibase 10.1016/S0167-2738(98)00483-4} {\bibfield  {journal}
		{\bibinfo  {journal} {Nature (Lodon)}\ }\textbf {\bibinfo {volume} {453}},\
		\bibinfo {pages} {80-83} (\bibinfo {year} {2008})}\BibitemShut {NoStop}%
	\bibitem [{\citenamefont {Stotland}\ and \citenamefont
		{Di Ventra}(2013)}]{stochasticmemory}%
	\BibitemOpen
	\bibfield  {author} {\bibinfo {author} {\bibfnamefont {A.}\ \bibnamefont
			{Stotland}} and \bibinfo {author} {\bibfnamefont {M.}~\bibnamefont {Di Ventra}} }\bibfield
	{title} {\enquote {\bibinfo {title} {Stocastic Memory: Memory enhancement due to noise},}\ }\href {\doibase 10.1103/PhysRevE.85.011116} {\bibfield  {journal}
		{\bibinfo  {journal} {Phys. Rev. E}\ }\textbf {\bibinfo {volume} {85}},\
		\bibinfo {pages} {011116} (\bibinfo {year} {2012})}\BibitemShut {NoStop}%
	\bibitem [{\citenamefont {Radecka}\ \emph {et~al.}(1999)\citenamefont
		{Radecka}, \citenamefont {Sobas},\ and\ \citenamefont
		{Rekas}}]{tio2}%
	\BibitemOpen
	\bibfield  {author} {\bibinfo {author} {\bibfnamefont {M.}\ \bibnamefont
			{Radecka}}, \bibinfo {author} {\bibfnamefont {P.}~\bibnamefont {Sobas}}, \ and\
		\bibinfo {author} {\bibfnamefont {M.}~\bibnamefont {Rekas}},\ }\bibfield
	{title} {\enquote {\bibinfo {title} {Ambipolar diffusion in TiO$_{2}$},}\ }\href {\doibase 10.1016/S0167-2738(98)00483-4} {\bibfield  {journal}
		{\bibinfo  {journal} {Solid State Ionics}\ }\textbf {\bibinfo {volume} {119}},\
		\bibinfo {pages} {55} (\bibinfo {year} {1999})}\BibitemShut {Stop}%

\end{thebibliography}
\end{document}